%% file: paper.tex
\documentclass[preprint2, times, twocolappendix]{aastex631}

\input{tex/build/macros/project.tex}
\input{tex/src/preamble-maneage.tex}

\input{tex/src/preamble-project.tex}

\begin{document}

\title{\projecttitle}

\author[0000-0003-1710-6613]{Mohammad Akhlaghi}
\affiliation{Centro de Estudios de Física del Cosmos de Aragón (CEFCA), Plaza San Juan 1, 44001, Teruel, Spain; \href{mailto:mohammad@akhlaghi.org}{mohammad@akhlaghi.org}}

\begin{abstract}
  \noindent
  The pointing pattern is an integral part of designing one's observation strategy for a certain scientific goal.
  But accounting for the particular science case or instrument artifacts (like distortion, vignetting or large areas of bad pixels) can make it hard to predict how the exposure map of the final stack will be.
  To help address this problem, Gnuastro 0.21 includes a new executable program called \texttt{astscript-pointing-simulate} that is fully described in the Gnuastro manual and comes with a complete tutorial.
  The figures of this research note are reproducible with Maneage, on the Git commit \projectversion.
\end{abstract}

\keywords{Astronomical methods (1043), Field of view (534), Direct imaging (387), Astronomical techniques (1684), Astronomy software (1855)}

\section{Introduction}\label{sec:intro}
\noindent
Astronomical images are often composed of many single exposures.
After completing the observations, when the science topic does not depend on the time of observation, we stack those single exposures into one deep \emph{stacked} image (also known as \emph{coadd}).
Designing the strategy to take those single exposures is therefore a very important aspect of planning an astronomical observation.

Each exposure has a \emph{pointing} (the location on the sky that it was targeting).
Subsequent exposures are usually taken with different pointings.
When the next pointing is ver near (a small fraction of the field of view) to the previous one, it is known as a \emph{dither} (literally/generally meaning trembling or vibration).
But when the distance between subsequent pointings is large (and issues like re-focusing become necessary), the pointing is known as an \emph{offset}.
These terms are sometimes used interchangably by some.

When we only have dithers, most of the central part of the final stack has a fixed depth.
Only a thin border becomes shallower (conveying a sense of vibration!).
For example see Figures 3 and 4 of \citet{xdf} which show the exposures that went into the XDF survey.
These types of images (where each pixel contains the number of exposures, or time, that were used in it) are known as exposure maps.
However, it can happen that only offsets are used.
For example see Figure 1 of \citet{lights}, which shows the exposure map for the LIGHTS survey.

\begin{figure*}[t]
  \ifdefined\makepdf%
    \tikzsetnextfilename{fig-pointing-demo}%
    \input{tex/src/fig-pointing-demo.tex}%
  \else
    \includegraphics[width=\linewidth]{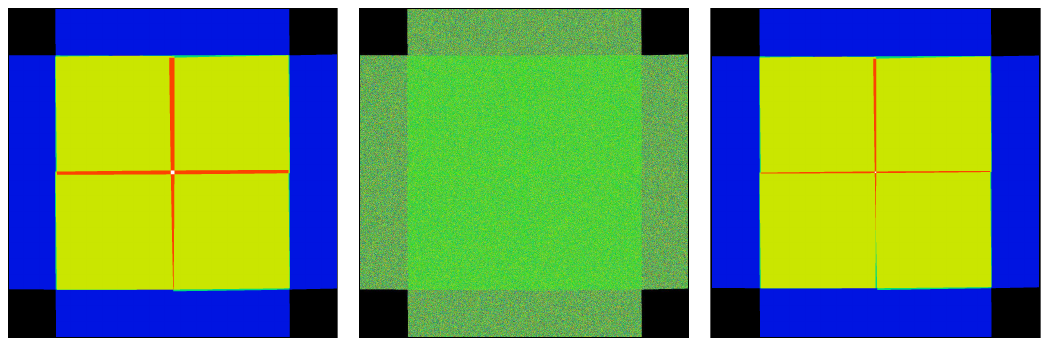}
  \fi

  \caption{\label{fig-pointing-demo}
    Sample outputs from the tutorial on designing a pointing pattern in the manual of Gnuastro 0.21.
    Left: the exposure map of a ``+''-shaped pointing (offset) pattern with 5 exposures.
    Center: The stack of pure noise after applying the same pointing list as the left image.
    Right: Exposure map of same dither pattern as the left image, but with trimming of the outer pixels in each exposure (this usually happens due to vignetting).
    The color map used shows the largest values as white/red and the smallest values as black/blue.
  }
\end{figure*}

The dithering pattern therefore is strongly defined by the science case (high-level purpose of the observation) and your telescope's field of view.
For example in the HUDF, CANDELS and other proposals (that constitute the XDF), the scientific target was high redshift (distant) galaxies.
When counting the pixels that they cover in relation to the number of pixels in each exposure, these targets are very small objects.
Such that within that small footprint (of just 1 arcminute!) we have thousands of such small objects (which may also be stars or nearby galaxies of similar observed size to the high redshift galaxies).
Therefore for the scientific goal of the XDF, the very small dithers were sufficient to avoid instrument artifacts.

However, the LIGHTS survey is focused on the halos of large nearby galaxies (that can be more than 10 arcminutes wide); thus covering a large fraction of the field of view of the Large Binocular Telescope that was used.
In order to do robust calibration of the images (for example to estimate the flat field pattern, or the sky background level) it was necessary to have offsets.
This enabled the galaxy and its halo to cover very different pixels from one exposure to the next.

In other cases, the pointings include both offsets and dithers (around each offset).
To find the ideal dither pattern for the particular scientific goal, it therefore helps to be able to simulate various strategies.
In the next section, an installed script is introduced for this purpose (to simulate the exposure map of a series of pointings) in GNU Astronomy Utilities \citep[Gnuastro][]{gnuastro} version\footnote{This paper is published shortly before the release of Gnuastro 0.21.
If Gnuastro 0.21 is not yet available, please use the latest test (alpha) release.} 0.21 and later.

\section{Simulating the exposure map}
Within Gnuastro, the executable in charge of simulating the exposure map of a given list of pointings is called \texttt{astscript\--pointing\--simulate}.
As with all Gnuastro features, it has a complete documentation in Gnuastro's manual\footnote{\url{https://www.gnu.org/software/gnuastro/manual}} which is available in many formats (on the command-line, as web pages, PDF and etc).
A complete and dedicated tutorial is also available within the tutorials chapter\footnote{\url{https://www.gnu.org/software/gnuastro/manual/html_node/Tutorials.html}} of the manual.

Three of the images produced in the tutorial (as of version 0.21) are displayed in Figure \ref{fig-pointing-demo}.
The left image shows the exposure map that is produced from a five-point, ``+''-shaped set of pointings on the sky, with one pointing in the center and four on the outer extents.
The middle image uses one of the \emph{hooks} that are available within the script to produce a noisy image and not an exposure map.
This hook can also be used to insert simulated/real objects.
The right image shows how another hook can be used to account for trimming of the outer edges (which sometimes do not get exposed enough due to vignetting).
This hook can also be used to insert any random area of bad pixels or to simulate persistence in near infra-red detectors (these can be large).

For its processing, this script needs two input files: 1) a table containing the RA and Dec of each pointing, 2) a FITS image from the camera that the dither pattern is meant for.
The image has to contain the celestial world coordinate system \citep[WCS, see][]{wcscelestial} headers.
Having a WCS is very important, especially for cameras with a wide field of view, because distortions can become visible/significant.

The tutorial and dedicated manual for this script are available in the Gnuastro manual and will always be up to date with the running version; therefore for the practical details, we encourage readers to follow the Gnuastro manual.
The future changes in the script that break with this research note will be clearly designated in a ``changes after publication'' subsection of the section describing this script.

\section{Acknowledgement}
The workflow of this research note was developed in the reproducible framework of Maneage \citep[\emph{Man}aging data lin\emph{eage},][latest Maneage commit \maneageversion{}, from \maneagedate]{maneage}.
This note is created from the Git commit {\projectversion} that is hosted on Codeberg\footnote{\url{\projectgitrepo}} and is archived on SoftwareHeritage for longevity.

The analysis of this research note was done using GNU Astronomy Utilities (Gnuastro, ascl.net/1801.009) version \gnuastroversion. Work on Gnuastro has been funded by the Japanese Ministry of Education, Culture, Sports, Science, and Technology (MEXT) scholarship and its Grant-in-Aid for Scientific Research (21244012, 24253003), the European Research Council (ERC) advanced grant 339659-MUSICOS, the Spanish Ministry of Economy and Competitiveness (MINECO, grant number AYA2016-76219-P) and the NextGenerationEU grant through the Recovery and Resilience Facility project ICTS-MRR-2021-03-CEFCA.

I also acknowledge the financial support provided by the Governments of Spain and Arag\'on through their general budgets and the Fondo de Inversiones de Teruel, and the Spanish Ministry of Science and Innovation (MCIN/AEI/10.13039/501100011033 y FEDER, Una manera de hacer Europa) with grant PID2021-124918NA-C43.

\bibliography{references}{}
\bibliographystyle{aasjournal}

\appendix
\section{Software acknowledgement}
\label{appendix:software}
\input{tex/build/macros/dependencies.tex}

\end{document}

%% file: tex/build/macros/project.tex
\input{tex/build/macros/initialize.tex}

\input{tex/build/macros/pointing-demo.tex}

\input{tex/build/macros/hardware-parameters.tex}

%% file: tex/build/macros/initialize.tex
\newcommand{\projecttitle}{Gnuastro: simulating the exposure map of a pointing pattern}
\newcommand{\projectversion}{4176d29}
\newcommand{\projectgitrepo}{https://codeberg.org/gnuastro/paper-pointing-simulate}
\newcommand{\projectcopyrightowner}{Mohammad Akhlaghi <mohammad@akhlaghi.org>}
\newcommand{\gnuastroversion}{0.20.72-08b0}
\newcommand{\maneagedate}{22 May 2023}
\newcommand{\maneageversion}{8161194}

%% file: tex/src/preamble-maneage.tex
\ifdefined\highlightnew

\else

\fi

\ifdefined\highlightnotes
\newcommand{\tonote}[1]{\textcolor{red!60!black}{[#1]}}
\else
\newcommand{\tonote}[1]{{}}
\fi

%% file: tex/src/preamble-project.tex
\usepackage{graphicx}

\usepackage{xcolor}
\color{black} 
\definecolor{DarkBlue}{RGB}{0,0,90}

\hypersetup{
    pdftitle={\projecttitle},
    pdfauthor={\projectcopyrightowner},
    pdfsubject={\projectgitrepo{} (commit \projectversion)},
    pdfkeywords={Astronomical methods, Field of view, Direct imaging, Astronomical techniques, Astronomy software, Reproducible research}
}

\input{tex/src/preamble-pgfplots.tex}

\shorttitle{\projecttitle}
\shortauthors{Akhlaghi, M}

%% file: tex/src/preamble-pgfplots.tex
\usepackage{tikz}
\usetikzlibrary{external}
\tikzsetexternalprefix{tex/tikz/}

\usepackage{pgfplots}
\pgfplotsset{compat=newest}
\usepgfplotslibrary{groupplots}
\pgfplotsset{
  axis line style={thick},
  tick style={semithick},
  tick label style = {font=\footnotesize},
  every axis label = {font=\footnotesize},
  legend style = {font=\footnotesize},
  label style = {font=\footnotesize}
  }

%% file: tex/src/fig-pointing-demo.tex
%
%
%
%

\begin{tikzpicture}
  \node[anchor=south] (img) at (0\linewidth,0\linewidth)
       {\includegraphics[width=0.31\linewidth]
         {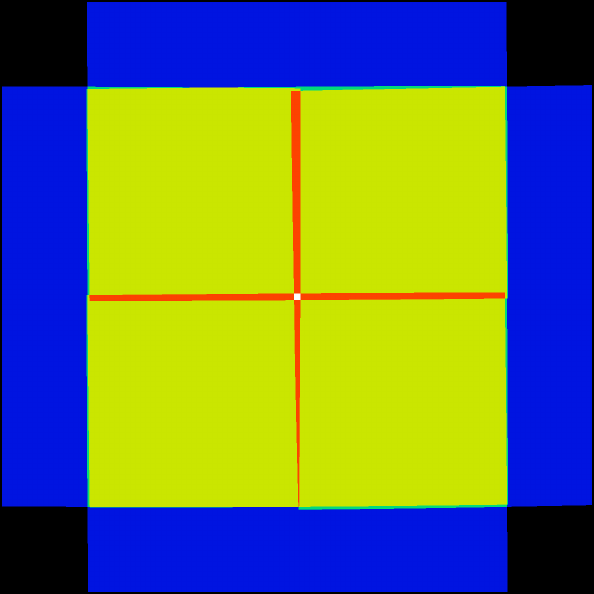}};

  \node[anchor=south] (img) at (0.33\linewidth,0\linewidth)
       {\includegraphics[width=0.31\linewidth]
         {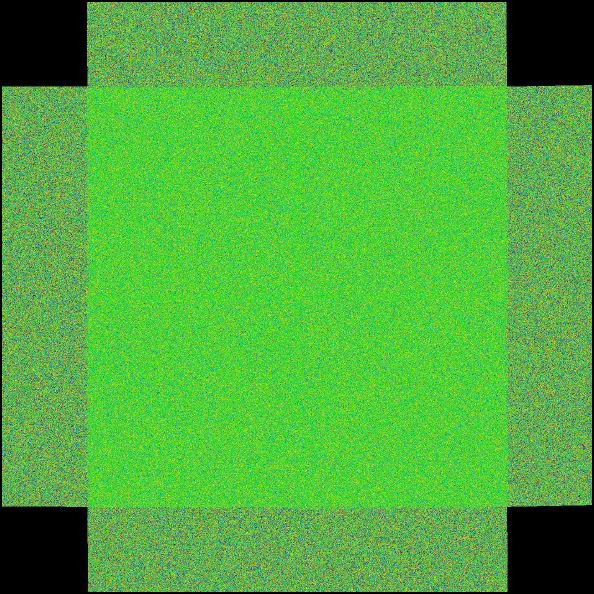}};

  \node[anchor=south] (img) at (0.66\linewidth,0\linewidth)
       {\includegraphics[width=0.31\linewidth]
         {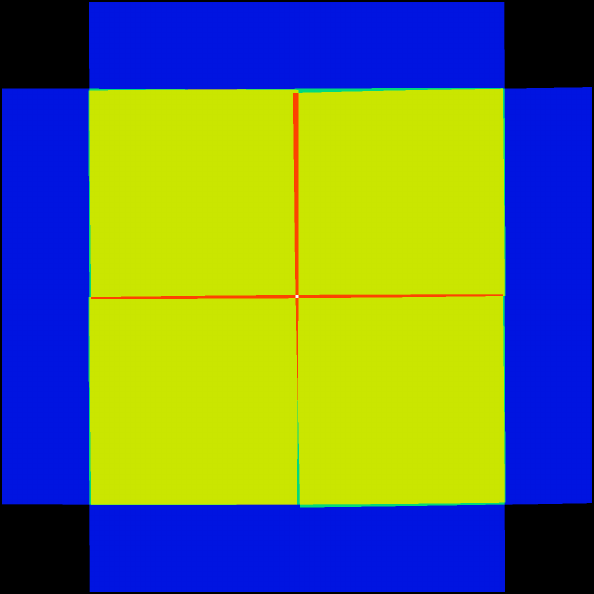}};
\end{tikzpicture}

%% file: tex/build/macros/dependencies.tex
 
This research was done with the following free software programs and libraries:  1.23, Bzip2 1.0.8, CFITSIO 4.1.0, CMake 3.24.0, cURL 7.84.0, Dash 0.5.11-057cd65, Discoteq flock 0.4.0, Expat 2.4.1, File 5.42, Fontconfig 2.14.0, FreeType 2.11.0, Git 2.37.1, GNU Astronomy Utilities 0.20.72-08b0 \citep{gnuastro,akhlaghi19}, GNU Autoconf 2.71, GNU Automake 1.16.5, GNU AWK 5.1.1, GNU Bash 5.2-rc2, GNU Binutils 2.39, GNU Bison 3.8.2, GNU Compiler Collection (GCC) 12.1.0, GNU Coreutils 9.1, GNU Diffutils 3.8, GNU Findutils 4.9.0, GNU gettext 0.21, GNU gperf 3.1, GNU Grep 3.7, GNU Gzip 1.12, GNU Integer Set Library 0.24, GNU libiconv 1.17, GNU Libtool 2.4.7, GNU libunistring 1.0, GNU M4 1.4.19, GNU Make 4.3, GNU Multiple Precision Arithmetic Library 6.2.1, GNU Multiple Precision Complex library, GNU Multiple Precision Floating-Point Reliably 4.1.0, GNU Nano 6.4, GNU NCURSES 6.3, GNU Readline 8.2-rc2, GNU Scientific Library 2.7, GNU Sed 4.8, GNU Tar 1.34, GNU Texinfo 6.8, GNU Wget 1.21.2, GNU Which 2.21, GPL Ghostscript 9.56.1, Help2man , Less 590, Libffi 3.4.2, Libgit2 1.3.0, libICE 1.0.10, Libidn 1.38, Libjpeg 9e, Libpaper 1.1.28, Libpng 1.6.37, libpthread-stubs (Xorg) 0.4, libSM 1.2.3, Libtiff 4.4.0, libXau (Xorg) 1.0.9, libxcb (Xorg) 1.15, libXdmcp (Xorg) 1.1.3, libXext 1.3.4, Libxml2 2.9.12, libXt 1.2.1, Lzip 1.23, OpenSSL 3.0.5, PatchELF 0.13, Perl 5.36.0, pkg-config 0.29.2, podlators 4.14, Python 3.10.6, util-Linux 2.38.1, util-macros (Xorg) 1.19.3, WCSLIB 7.11, X11 library 1.8, XCB-proto (Xorg) 1.15, xorgproto 2022.1, xtrans (Xorg) 1.4.0, XZ Utils 5.2.5 and Zlib 1.2.11. 
The \LaTeX{} source of the paper was compiled to make the PDF using the following packages: courier 61719 (revision), epsf 2.7.4, etoolbox 2.5k, helvetic 61719 (revision), lineno 5.3, pgf 3.1.10, pgfplots 1.18.1, revtex4-1 4.1s, tex 3.141592653, textcase 1.04 and ulem 53365 (revision). 
We are very grateful to all their creators for freely  providing this necessary infrastructure. This research  (and many other projects) would not be possible without  them.